\documentclass[11pt,a4paper]{article}

\usepackage[utf8]{inputenc} 
\usepackage[T1]{fontenc}    %
\usepackage[english]{babel} 
\usepackage[final]{microtype} 

\usepackage{amssymb,amsmath,mathtools} 
\usepackage{amsthm} 

\usepackage{xcolor}
\usepackage{bbm}
\usepackage{braket}

\usepackage[hmargin=0.12\paperwidth,vmargin=0.16\paperwidth,bindingoffset=0cm]%
{geometry} 

\numberwithin{equation}{section} 

\theoremstyle{plain}

\theoremstyle{definition}

  \numberwithin{prop}{section}
   \numberwithin{cor}{section}
   \numberwithin{remark}{section}

\newcommand{\MeijerG}[8][\Big]{G^{{ #2 },{ #3 }}_{{ #4 },{ #5 }} #1( \begin{matrix} #6 \\ #7 \end{matrix}\, #1\vert\, #8 #1)}

\title{\Large\bfseries In memoriam: aspects of Santosh Kumar's work on exact results in RMT}%
   
\author{Peter J. Forrester}
\date{}

\begin{document}

\maketitle

School of Mathematics and Statistics,  The University of Melbourne,
Victoria 3010, Australia. \: \: Email: {\tt pjforr@unimelb.edu.au}; \\

\bigskip

\begin{abstract}
\noindent
Santosh Kumar was an active researcher on the topic of exact results in random matrix theory and their various applications, particularly to
quantum chaos and information theory. Barely entering his mid-career, he died unexpectedly on the 18th October 2024. The present article
gives an account of some of his research directions and findings. As well as serving as a tribute to his work, this is done also for the
purpose of providing a resource for those who may continue along related  lines in the future.
 
\end{abstract}

  \setcounter{section}{-1}
 \section{Prologue}
 
 Tragedy struck for the family, friends, students and colleagues of Santosh Kumar in October last year (2024) with his sudden and unexpected death.
 At the time, he was working alongside, or had just graduated, several PhD students, whereby he was overseeing a steady stream of publications 
 relating to exact results in random matrix theory (RMT).  Typically these works were targeted to, or motivated by, applied domains, with
 quantum information, quantum chaos, telecommunications and complex systems all prominent.
 In addition, he was carrying out multiple lecturing and administrative duties at his home university, the Shiv Nadar Institution of Eminence, including supervising
 undergraduate students to provide them with a research experience. He was also collaborating internationally, with a further paper in a fruitful series of works
 with the present author  relating to
 the exact computation of random matrix distribution for the classical random matrix ensembles in preparation at the time of his untimely passing.
 
 By way of respect and appreciation for the benefit of a long standing collaboration (began in 2016, with 5 published works, one paper still in the refereeing process, and one
 preprint), it seems only fitting to document some of Santosh Kumar's research finding by way of a review of aspects of his work. To give this context, some background material will also
 be provided. This review is ordered by various common themes where can be found in several of his works. However, as I personally don't have expertise across them all, some will not be represented. Notable among these are his research applying RMT to telecommunications (a selection of such works are \cite{KP10,SKK17,CKFM18,DJGK24}) and his
 work as a postdoc relating to the use of supersymmetry methods to analyse scattering matrix elements \cite{K+13,NKSG14}. 
 One should draw attention to his emerging interest on aspects of non-Hermitian random matrix theory \cite{SSK23a,SK24}, which otherwise is not discussed.
 Also, it will not be possible to do justice to the careful
 numerical validation of theoretical predictions, often making use of the computer algebra software Mathematica, nor the demonstration of their applicability to model physical systems (in quantum chaos studies, the kicked top features prominently \cite{KSA17,SK21}, as do spin chains \cite{SK23,SSK23}).

\section{Quantifying entanglement}\label{S2}
\subsection{Formalism and relationship to RMT}
Analytic formulas relating to a particular class of the
quantum entanglement with a random matrix origin have been known in the literature for several decades now, beginning it would
seem with \cite{Lu78,LP88,Pa93}. The physical setting relates to the bipartite entanglement of two finite dimensional subsystems, $A$ and
$B$ say. Let their respective dimensions be $n$ and $N$, with $\{ |a_i \rangle_{i=1,\dots,n} \}$, $\{ |b_i \rangle_{i=1,\dots,n} \}_{i=1,\dots,N}$
and take $n \ge N$ for definiteness. A general state $| \psi \rangle$ in the composite system can be written as a linear combination of the
$nN$ dimensional tensor product basis $\{ | a_i \rangle \otimes | b_j \rangle \}$,
\begin{equation}\label{AD1x}
| \psi \rangle_{AB} = \sum_{i=1}^n \sum_{j=1}^N x_{ij}  | a_i \rangle \otimes | b_j \rangle,
\end{equation}
where normalisation requires 
\begin{equation}\label{AD2}
\sum_{i=1}^n \sum_{j=1}^N | x_{ij} |^2 = 1.
\end{equation}

Associate with (\ref{AD1x}) the rectangular matrix $X=[x_{ij}]_{i=1,\dots,n \atop j=1,\dots,N}$. 
In the circumstance that ${\rm rank} \, X = 1$, there are bases for the subsystems $A,B$ such that (\ref{AD1}) decomposes as
as a single tensor product of a state from subsystem $A$, and a state from subsystem $B$. In this circumstance (\ref{AD1x}) is said
to be separable --- otherwise it is entangled.
A classical result from the theory of matrices
gives the so-called singular value decomposition
\begin{equation}\label{SV1} 
X = U L V^\dagger,
\end{equation}
where $U$ is of size $n \times n$, $L$ is of size $n \times N$ and $V$ is of size $N \times N$.
In (\ref{SV1}), the entries at diagonal positions $(ii)$ ($i=1,\dots,N$) of $L$ are the square root of the eigenvalues of
$X^\dagger X$, $\{ \sqrt{\lambda_i} \}$ say, referred to as the singular values.  According to the normalisation condition (\ref{AD2}) written as
${\rm Tr} \, X^\dagger X = 1$, a constraint on these eigenvalues (as well as being positive) is that $\sum_{i=1}^N \lambda_i = 1$.
All other entries of this matrix are zero. The rows of $V^\dagger$ are an orthonormal basis for ${\rm span} \, \{ | b_j \rangle \}_{j=1,\dots,N}$,
while the columns of $\{ |a_i   \rangle \}_{i=1,\dots,n}$ are a orthonormal basis for $U$. Let these new orthonormal bases be specified as
$\{ | v_j^B \rangle \}_{j=1,\dots,N}$ and $\{ | u_j^A \rangle \}_{j=1,\dots,n}$. They allow (\ref{AD2}) to be simplified,
\begin{equation}\label{AD1}
| \psi \rangle_{AB} =   \sum_{i=1}^N  \sqrt{\lambda_i} | u_i^A \rangle \otimes | v_i^B \rangle,
\end{equation}
giving what is referred to as the Schmidt decomposition (see e.g.~\cite{ES20} in relation to the naming).

By definition, the density matrix associated with the state (\ref{AD1x}) of the composite system is the $n N \times n N$ rank one matrix
\begin{equation}\label{rAB} 
\rho_{AB} = | \psi \rangle_{AB}  \,  {}_{AB}\langle \psi |.
\end{equation}
The reduced density matrix of subsystem $A$ obtained by tracing out over subsystem $B$,\footnote{Here there is an abuse of notation; one requires
that $\langle a_i | (|a_{i'} \rangle \otimes | b_{j} \rangle)  := \delta_{i,i'}  | b_{j} \rangle$}
\begin{equation}\label{rAB1} 
\rho_A : = \sum_{i=1}^N \langle a_i |  \rho_{AB} | a_i \rangle.
\end{equation}
In terms of the Schmidt decomposition (\ref{AD1}) this takes the form
\begin{equation}\label{rAB2} 
\rho_A = \sum_{i=1}^N \lambda_i  | u_i^A \rangle \langle u_i^A |,
\end{equation}
which one recognises as the form of a density matrix for a statistical mixture of pure states, each state $| u_i^A \rangle$ occurring with probability $p_i$
(see e.g. the recent review \cite{B+22}). Moreover, from the meaning of the quantities in (\ref{AD1}), one has the explicit matrix form
\begin{equation}\label{rAB3a} 
\rho_A = X^\dagger X.
\end{equation}

Having arrived at the formula (\ref{rAB3a}), one would like to introduce randomness into the problem by devising an ensemble theory. This would be in keeping
with the coefficients in (\ref{AD1x}) making up the matrix $X$ being random, up to the constraint (\ref{AD2}). Actually the requirement that
$X$ (in distribution) is unchanged by left and right conjugation by unitary matrices should also  be added as a constraint, as only then will $\rho_A$ be
independent of the particular choice of basis in (\ref{AD1x}). With an eye towards exact solutions, and knowing the special role played in random matrix theory by
the Gaussian distribution from this viewpoint (see e.g.~\cite{Me04,Fo10}, one then is lead to the choice
\begin{equation}\label{XT}
X = G/{\rm Tr}(G^\dagger G)^{1/2},
\end{equation}
where 
the $n \times N$ matrix $G$ has
independent standard complex Gaussian entries (an example of a rectangular GinUE matrix \cite{BF25}).
The matrix $G$ has
 the bi-unitary invariant distribution proportional to
$e^{- {\rm Tr} \, G^\dagger G}$. A normalisation by ${\rm Tr} \, (G^\dagger G)^{1/2}$ has been carried out so that $X$ then obeys the constraint
 (\ref{AD2}). Hence the distribution on the space of matrices $X$ so constructed is proportional to $\delta(1 - {\rm Tr}(G^\dagger G))$.
Further, in consideration of a (rectangular) matrix $X$ random with complex elements, it is a standard result in random matrix theory
(see e.g.~\cite[Eq.~(3.23)]{Fo10}) that changing variables to $W = X^\dagger X$ gives a Jacobian factor $(\det W)^{n-N}$, so the
distribution on $W$ is proportional to 
\begin{equation}\label{W1}
(\det W)^{n-N} \delta(1 - {\rm Tr} \, W). 
\end{equation}
Another standard result is that changing variables
from the elements of $W$ to its eigenvalues and eigenvectors gives a Jacobian factor $\prod_{1 \le j < k \le N} (\lambda_k - \lambda_j)^2$
\cite[Eq.~(1.27) with $\beta = 2$]{Fo10}. This can be deduced from the metric on Hermitian matrices corresponding to the infinitesimal squared 
line element $(ds)^2 = {\rm Tr} (dW)^2$ \cite[\S 1.2.4]{Fo10}, which in turn corresponds to the distance function between Hermitian matrices
\begin{equation}\label{W1a}
(d^{\rm HS}(A,B))^2 = {\rm Tr} \, (A - B)^2,
\end{equation}
where the superscript "HS" stands for Hilbert-Schmidt. 
Hence one is led \cite{Ha98,ZS03} to a probability density function (PDF)
on the $\{ \lambda_j \}$ in (\ref{rAB2})
proportional to
\begin{equation}\label{rAB3} 
\delta \Big ( 1 - \sum_{l=1}^N \lambda_l \Big ) \prod_{l=1}^N \lambda_l^{n-N} \mathbbm 1_{\lambda_l > 0} \prod_{1 \le j < k \le N} (\lambda_k - \lambda_j)^2,
\end{equation}
which due to its relation to (\ref{W1a}) is referred to as the Hilbert-Schmidt measure for density matrices.

For $A,B$ density matrices, an alternative to (\ref{W1a}) is the distance function
\begin{equation}\label{W1b}
(d^{\rm B}(A,B))^2 =  2 - 2\, {\rm Tr} \, (A^{1/2} B A^{1/2})^{1/2},
\end{equation}
which enjoys several distinguished properties \cite{BZ06}. Forming a line element of a single density matrix based on this Bures distance function, it was shown by Hall
\cite{Ha98}
that the implied Jacobian for a change of variables to the eigenvalues and eigenvectors gives the Jacobian
$\prod_{l=1}^N \lambda_l^{-1/2} \prod_{1 \le j < k \le N} (\lambda_k - \lambda_j)/(\lambda_k + \lambda_j)$. This then gives
rise to the PDF for $\{\lambda_l\}$
proportional to
\begin{equation}\label{rAB4} 
\delta \Big ( 1 - \sum_{l=1}^N \lambda_l \Big ) \prod_{l=1}^N  \lambda_l^{n-N-1/2} \mathbbm 1_{\lambda_l > 0} \prod_{1 \le j < k \le N} { (\lambda_k - \lambda_j)^2 \over
 \lambda_k + \lambda_j },
\end{equation}
 referred to as the Bures-Hall (BH) measure for density matrices. Replacing $G$ in (\ref{XT}) by $G(\mathbb I + U)$, where $U$ is a random unitary matrix chosen with PDF
proportional to $| \det (\mathbb I + U) |^{2(n-N)}$ gives a realisation of random density matrices with eigenvalue PDF (\ref{rAB4}) \cite{OSZ10}. 

Scaling $\lambda_l \mapsto \lambda_l/t$ $(t > 0)$ in (\ref{rAB3}) and (\ref{rAB4}) and factoring out powers of $t$ where possible shows that the functional forms are unchanged except that $\delta ( 1 - \sum_{l=1}^N \lambda_l) \mapsto \delta ( t - \sum_{l=1}^N \lambda_l)$. Now integrating over this variable to form the Laplace transform then replaces this delta function term by the exponential $e^{-\sum_{l=1}^N x_l}$ (the Laplace transform variable can be set equal to unity by a further rescaling). The Hilbert-Schmidt measure (\ref{rAB3}) then becomes the eigenvalue PDF specifying the complex Wishart ensemble (or Laguerre unitary ensemble by regarding $n-N = a$ as a continuous variable), while the transformed form of (\ref{rAB4}) may be referred to as the Laguerre  Bures-Hall  ensemble. In exact calculations relating to (\ref{rAB3}) and (\ref{rAB4}), it is often the case that the quantity of interest can first be deduced for its Laguerre counterpart, and then an inverse Laplace transform performed. For example, if $\rho_{(1),N}^{\rm HS}(x)$ denotes the eigenvalue density for the Hilbert-Schmidt measure, and  $\rho_{(1),N}^{\rm LUE}(x)$ denotes the eigenvalue density for the Laguerre unitary ensemble, one has
\begin{equation}\label{rAB5} 
{\rho}_{(1),N}^{\rm HS}(x)  =   { C_{N,a}^{\rm LUE} \over C_{N,a}^{\rm HS} } {1 \over 2 \pi i} \int_{c - i \infty}^{c + i \infty}
{  {\rho}_{(1),N}^{\rm  LUE} (sx) \over s^{N a + N^2 - 1}} e^s \, ds, \quad c > 0,
\end{equation}
where $C_{N,a}^{\rm LUE},  C_{N,a}^{\rm HS}$ are the normalisations; see e.g.~\cite{ATK11}.

\subsection{Some results of Kumar}
\subsubsection{Entanglement statistics}
Statistical quantities often used to quantify entanglement of a bipartite system as encoded in (\ref{rAB2})
are von Neumann entropy $\mu_{\rm vN}$ and purity $\mu_2$ specified by
\begin{equation}\label{rAB6} 
\mu_{\rm vN} := \sum_{j=1}^N \lambda_j \log \lambda_j, \qquad \mu_2 := \sum_{j=1}^N \lambda_j^2.
\end{equation}
Let $\psi(x)$ denote the digamma function. In the Hilbert-Schmidt case it has been known for a long time 
\cite{Pa93,Se96,Gi07a,SZ04} that
\begin{equation}\label{rAB7} 
\mathbb E_{\rm HS} ( \mu_{\rm vN} ) = \psi(nN+1) - \psi(n+1) - {N - 1 \over 2 n}, \quad \mathbb E_{\rm HS} (\mu_2) = {n+N \over n N + 1}.
\end{equation}
By way of interpretation, one notes that in the limit $n \to \infty$ (in the bipartite formalism, $n$ is the dimension of subsystem $B$), these formulas
give $\mathbb E_{\rm HS}  ( \mu_{\rm vN}) \to \log N$, $\mathbb E_{\rm HS}  ( \mu_2) \to 1/N$, which is to be expected as then $\rho_A$ approaches ${1 \over N} \mathbb I_N$,
as can be from its construction (\ref{rAB7}). Also special is the case $n=N=1$ corresponding to a pure state so that $\rho_A$ has a single eigenvalue,
which equals unity, and thus $\mu_{\rm vN} =0$, $ \mu_2 =1$, which indeed are properties of (\ref{rAB7}).

In \cite{SK19}, Sarkar and Kumar took up the problem of computing $\mu_{\rm vN}$ and  $\mu_2$ for the Bures-Hall measure. Their approach was
 to use a known Pfaffian structure associated with the Laguerre  Bures-Hall  ensemble \cite{FK16} to obtain a computable formula for the one-point density, which was then lifted to a computable formula for the one-point density of the  Bures-Hall measure using (an appropriate modification of) (\ref{rAB5}).
Noting that both quantities in (\ref{rAB6}) have the structure $\sum_{l=1}^N f(\lambda_l)$ of a linear statistic, use was then made of the general formula
\begin{equation}\label{M1}
\mathbb E \Big ( \sum_{l=1}^N f(\lambda_l) \Big ) = \int_0^1 f(\lambda) \rho_{(1),N}(\lambda) \, d \lambda
\end{equation}
(here the fact that the eigenvalues are supported on $(0,1)$ has been used in the terminals of integration) to obtain the exact (rational) values of the means. Moreover, it was found
that the tabulated values were consistent with the functional forms
\begin{equation}\label{rAB8} 
\mathbb E_{\rm BH} ( \mu_{\rm vN} ) = \psi(nN-N^2/2+1) - \psi(n+1/2), \quad \mathbb E_{\rm BH} (\mu_2) = {2n(2n+N) + (N^2 - 1) \over 2 n (2n N - N^2 + 2)};
\end{equation}
one can verify (as was done in \cite{SK19}) that the consistency checks noted below (\ref{rAB7}) are features of these formulas. For $n=N \to \infty$ the first of these gives
$\mathbb E_{\rm BH} ( \mu_{\rm vN} )  \sim \log N - \log 2$, whereas (\ref{rAB6}) gives $\mathbb E_{\rm HS} ( \mu_{\rm vN} )  \sim \log N - {1 \over 2}$; see also \cite{SZ04}.

The conjectures (\ref{rAB8}) inspired multiple works by Wei \cite{We20a,We20b}, and Wei and collaborators \cite{LW21,WW23,YHOW24}, relating to measures of entanglement with respect to the Bures-Hall measure. In particular, in \cite{We20a} a proof was given of (\ref{rAB8}). A key ingredient for this was to supplement the Pfaffian structure as already used in \cite{SK19}, with the explicit functional form for the eigenvalue density in terms of Meijer $G$-functions known from \cite{FK16}. Special function properties of the Meijer $G$-function were then used to facilitate the evaluation of the averages as specified by (\ref{M1}). Generally, in relation to the statistical properties of a linear statistic, at the next level of complexity beyond the mean is the variance. This is given in terms of the truncated (or connected) two-point correlation $\rho_{(2),N}^T$ according to
\begin{equation}\label{d1}
{\rm Var} \, \Big ( \sum_{j=1}^N f(\lambda_j) \Big ) = \int_0^1 d \lambda_1 \, f(\lambda_1)  \int_0^1 d \lambda_2 \, f(\lambda_2) \Big (
\rho_{(2),N}^T(\lambda_1,\lambda_2) + \delta(\lambda_1 - \lambda_2) \rho_{(1),N}^T(\lambda_1) \Big );
\end{equation}
see e.g.~\cite[Prop.~2.1]{Fo23}. Beginning with (\ref{d1}), and knowledge of $\rho_{(2),N}^T$ from \cite{FK16}, Wei \cite{We20b} extended the exact result for
$\mathbb E_{\rm BH} ( \mu_{\rm vN} )$ in (\ref{rAB8}) by deriving that
\begin{equation}\label{d2}
{\rm Var}_{\rm BH}( \mu_{\rm vN} ) = - \psi_1(nN-N^2/2+1) + {2n(2n+N) - N^2 +1 \over 2n (2nN - N^2 + 2)} \psi_1(n+1/2),
\end{equation}
where $\psi_1(\ell) := {\pi^2 \over 6} - \sum_{k=1}^{\ell - 1} {1 \over k^2}$ is the trigamma function. Since $\psi_1(\ell) \sim {1 \over \ell}$ as $\ell \to \infty$, the large $N$ asymptotic
form of (\ref{d2}) is, for $n/N \to c \ge 1$, ${\rm Var}_{\rm BH}( \mu_{\rm vN} ) \sim {1 \over 2 c^2 N^2}$. A conjecture from \cite{We20b} is the asymptotic Gaussian fluctuation formula
\begin{equation}\label{d3}
{ \mu_{\rm vN} - \mathbb E_{\rm BH} ( \mu_{\rm vN}) \over ( {\rm Var}_{\rm BH}( \mu_{\rm vN} ) )^{1/2}} \to {\rm N}[0,1].
\end{equation}

\subsubsection{Density matrix distances}
The distance functions underpinning the Hilbert-Schmidt measure, and the Bures-Hall measure, are (\ref{W1a}) and (\ref{W1b}) respectively. Construction of the density matrices
for each of these measures in the form $X^\dagger X$, with $X$ specified in terms of  (rectangular) standard complex Gaussian matrices has been noted in
(\ref{XT}) (Hilbert-Schmidt case) and in the text below (\ref{rAB4}) (Bures-Hall case). In the work \cite{Ku20}, Kumar initiated the study of the distance function  (\ref{W1a}),
with $A,B$ random density matrices as relate to the Hilbert-Schmidt measure, with one of the density matrices relating to a rectangular matrix of $X$ of size $n_1 \times N$, and the other to a rectangular matrix of $X$ of size $n_2 \times N$. In this setting it was shown 
\begin{equation}\label{e1}
\mathbb E_{\rm HS}  {\rm Tr}(A - B)^2 = {N + n_1 \over N n_1 + 1} + {N + n_2 \over N n_2 + 1} - {2  \over N} \sim {1 \over N} {c_1 + c_2 \over c_1 c_2},
\end{equation}
where $c_1 = n_1/N$ and $c_2 = n_2/N$; see also \cite{PPZ16} in relation to the asymptotic form for $c_1 = c_2 = 1$. In a subsequent work \cite{LK24}, Laha and Kumar extended
this exact result for the mean to an exact evaulation of the corresponding variance. We make note only of its large $N$ asymptotic form
\begin{equation}\label{e1a}
{\rm Var}_{\rm HS}  {\rm Tr}(A - B)^2 \sim   {2 (c_1 + c_2)^2 \over c_1^2 c_2^2 N^4}.
\end{equation}
The works \cite{Ku20} and \cite{LK24} also contain exact results in the case that only one of the density matrices $A,B$ is random from the Hilbert-Schmidt measure, with the other density matrix fixed.

In (\ref{W1b}) the quantity
\begin{equation}\label{e2}
( \mathcal F(A,B) )^{1/2} := {\rm Tr} (A^{1/2} B A^{1/2})^{1/2} = \sum_{j=1}^N \lambda_j^{1/2}
\end{equation}
is referred to as the (square root of) the fidelity. Here $\{ \lambda_j \}$ are the eigenvalues of the matrix product $A^{1/2} B A^{1/2}$.
In the case that $A,B$ are complex Wishart matrices, and thus relate to density matrices with Hilbert-Schmidt measure but without the fixed trace constraint, in
\cite{AIK13} the statistical state formed by the eigenvalues of such matrix products have been analysed as examples of specific biorthogonal ensembles
\cite{Bo98} involving Meijer $G$-functions. From this starting point, Laha, Aggarwal and Kumar \cite{LAK21}, and Laha and Kumar \cite{LK23}
computed for the average square root fidelity
\begin{equation}\label{e3}
\mathbb E_{\rm HS}  \mathcal (F(A,B) )^{1/2}  = {2 \over \pi (N n_1)_{1/2} (N n_2)_{1/2}}
\sum_{j=1}^N (j)_{1/2} (j + n_1 - N)_{1/2} (j + n_2 - N)_{1/2} (N - j + 1)_{-1/2},
\end{equation}
where $(u)_a := \Gamma(u+a)/\Gamma(u)$, and for the average fidelity
\begin{eqnarray}\label{e4}
\mathbb E_{\rm HS}  \mathcal F(A,B)   = {1 \over N} + {8 \over \pi^2 N^2 n_1 n_2}
\sum_{1 \le j < k \le N} { (j-k)^2 \over (j-k)^2 - 1/4} (j)_{1/2} (j+ v_1)_{1/2} \nonumber \\
\times  (j+ v_2)_{1/2} (N - j + 1)_{-1/2} (k)_{1/2} (k+ v_1)_{1/2} 
  (k+ v_2)_{1/2} (N - k + 1)_{-1/2} ,
\end{eqnarray}
where $v_1 = n_1 - N$ and $v_2 = n_2 - N$.

\subsection{Statistics of the smallest eigenvalue}\label{S1.3}
In \cite{KSA17}, Kumar and his students pointed out that the unit trace requirement implies the constraint $0 \le \lambda_N \le 1/N$ for
the smallest eigenvalue. Significantly, $\lambda_N \approx 1/N$ implies that all the remaining $(N-1)$ eigenvalues must similarly take a value
approximately equal to $1/N$, which is the condition for (close to) maximum von Neumann entropy. Further, the other extreme $\lambda_N \approx 0$ can
be taken as an indicator that the effective Hilbert space dimension of the subsystem under consideration can be effectively reduced by one.
These observations motivated a study of the exact computation of the distribution of the smallest eigenvalue for the Hilbert-Schmidt measure.

Consider first the complex Wishart ensemble, which replaces the delta function in (\ref{rAB3}) by an exponential. The probability of no eigenvalues in
$(0,s)$ is, with $a:= n - N$
\begin{align}\label{e5}
E_N(s;a) & := {1 \over C_{N,a}} \int_s^\infty dx_1 \cdots  \int_s^\infty dx_N \, \prod_{l=1}^N x_l^a e^{- x_l} \prod_{1 \le j < k \le N} (x_k - x_j)^2 \nonumber \\
& = {e^{-Ns} \over C_{N,a}} \int_0^\infty dx_1 \cdots  \int_0^\infty dx_N \, \prod_{l=1}^N (x_l+s)^a e^{- x_l} \prod_{1 \le j < k \le N} (x_k - x_j)^2,
\end{align}
where $C_{N,a}$ denotes the normalisation.
With $a$ a non-negative integer, one sees that the multi-dimensional integral is a polynomial of degree $Na$ in $s$ and thus
\begin{equation}\label{e5a}
E_N(s;a)  = e^{-Ns}  \sum_{l=0}^{Na} c_{l} s^l
\end{equation}
for some coefficients $\{c_l\}$. Now let $f_N(s;a)$ denote the PDF for the smallest eigenvalue. One has the general relation
$f_N(s;a) = - {d \over ds} E_N(s;a) $, which combined with the fact that $f_N(s;a)$ must be proportional to $s^a$ for small $s$
gives that $f_N(s;a) =  e^{-Ns}  \sum_{l=a+1}^{Na+1} d_l s^{l-1}$ for some coefficients  $\{d_l\}$ simply related to $\{c_l\}$. 
Now let $f_N^{\rm HS}(s;a)$ denote the smallest eigenvalue PDF for the Hilbert-Schmidt measure. By use of the inverse Laplace transform
(recall (\ref{rAB5}) and associated text)  it then follows
\begin{equation}\label{e5b}
f_N^{\rm HS}(s;a) = \Gamma(nN) \sum_{j=a+1}^{aN+1}  d_l  {(1 - n s)^{N(N+a) - j - 1} s^{j-1} \over \Gamma(N(N+a) - j)}
\Theta(1 - Ns),
\end{equation}
where $\Theta(u)$ denotes the Heaviside function. Two possible ways to compute $\{d_l\}$ are noted in \cite{KSA17}. One is to make use of
a known \cite{FH94} determinant formula for the polynomial portion of $f_N(s;a)$, involving an $a \times a$ Wronskian type matrix based on certain Laguerre
polynomials. The other is to make use of a modification of a differential recursion scheme from the theory of Selberg integrals \cite{Ao87}, \cite{Ed88}, \cite[\S 4.6]{Fo10},
well suited for implementation in computer algebra. With this carried out, (\ref{e5b}) gives the exact functional form of $f_N^{\rm HS}(s;a)$ from which
statistical properties of interest, for example the mean position of the smallest eigenvalue, can similarly be computed exactly.

\section{Real eigenvalues of certain integrable asymmetric random matrix ensembles and the Meijer $G$-function}
\subsection{A conjectured arithmetic property}
The Meijer $G$-function, which has already been mentioned in passing in \S \ref{S2}, is defined as the
particular Mellin transform
\begin{equation}\label{MG}
\MeijerG{m}{n}{p}{q}{a_1,\ldots,a_p}{b_1,\ldots,b_q }{z}=\frac{1}{2\pi i}\int_\gamma
\frac{\prod_{j=1}^m\Gamma(b_j+u)\prod_{j=1}^n\Gamma(1-a_j-u)}
{\prod_{j=m+1}^q\Gamma(1-b_j-u)\prod_{j=n+1}^p\Gamma(a_j+u)}z^{-u}
\, du.
\end{equation}
Here the contour $\gamma$ goes from $-i \infty$ to $i \infty$ with the poles of $\Gamma(b_j - s)$ on the left and those of $\Gamma(1 - a_j + s)$ on the right;
see \cite{Lu69} in relation to its many special properties.

Our interest in this section on the Meijer $G$-function is its appearance in a formula for the probability,
$p_{m,N}$ say, that all the eigenvalues in the product of $m$ independent $N \times N$ standard real Gaussian matrices
(also referred to as GinOE matrices \cite{BF25}). 
The result depends on the parity of $N$. With $N$ even for definiteness, we have from
\cite{Fo13} that
\begin{equation}\label{11a}
p_{m,N} = \Big ( \prod_{j=1}^N {1 \over \Gamma (j/2) } \Big )^m
\det \Big [  G_{m+1,m+1}^{m+1,m} \Big (   {5/2-j,\dots, 5/2-j,2 \atop 1, 1+k,\dots, 1+k} \Big | 1 \Big ) \Big ]_{j,k=1,\dots,N/2}
\end{equation}
Specialising now to the case $m=2$ (i.e.~the product of two standard real Gaussian matrices), evidence (by way of
high precision numerical evaluation) that all the entries of (\ref{11a}) are of the form $\pi^2$ times a rational number,
which was put forward as a conjecture. Note that the arithmetic structure of the probability of $p_{m,N} |_{m=2}$ itself
as a power of $\pi$ times a rational number then follows as a corollary. Earlier, in \cite{Ed97}, it had been shown that
$p_{m,N} |_{m=1}$ is equal to the rational number $2^{-N(N-1)/2}$, and special arithmetic properties of the probability of some specific
number $k$ of real eigenvalues for
the product of a standard real Gaussian matrix
and the inverse of a standard real Gaussian matrix had been deduced \cite{FM11}.
\subsection{Kumar's proof}
The work of Kumar \cite{Ku15} gave a proof of the above conjecture, by deriving a closed form expression for the particular
$ G_{3,3}^{3,2}$ as finite sums. At the same time this provides an efficient and exact computational scheme
for the entries in (\ref{11a}). 

The methodology behind such formulas, presented in more generality in the subsequent work
\cite{FIK20}, is to combined the eneral three-term recurrence relation for Meijer $G$-functions~\cite{Lu69}
\begin{equation}\label{3.3}
\MeijerG{m}{n}{p}{q}{a_1,\ldots,a_p}{b_1,\ldots,b_q}{z}=
\frac{\MeijerG{m}{n}{p}{q}{a_1,\ldots,a_{p-1},a_p-1}{b_1,\ldots,b_q}{z}+
\MeijerG{m}{n}{p}{q}{a_1,\ldots,a_p}{b_1,\ldots,b_{q-1},b_q+1}{z}}{a_p-b_q-1}
\end{equation}
for $n<p$ and $m<q$; together with evaluations~\cite{Lu69,FK16}
\begin{align}
\MeijerG{m+1}{m}{2m+1}{2m+1}{\frac32-j,\ldots,\frac32-j;\ell_1+k,\ldots,\ell_m+k,1}{0,k,\ldots,k;\frac32-j,\ldots,\frac32-j}{1}&=0, \label{3.4}\\
\MeijerG{m+1}{m}{2m+1}{2m+1}{\frac32-j,\ldots,\frac32-j;k,\ldots,k,1}{0,k,\ldots,k;\frac32-j-\ell_1,\ldots,\frac32-j-\ell_m}{1}
&=\prod_{i=1}^m\frac{\Gamma(j-\frac12)}{\Gamma(j-\frac12+\ell_i)} \label{3.5}
\end{align}
for non-negative integers $\ell_1,\ldots,\ell_m$, when not all of them are $0$. 

Through the use of (\ref{3.3})--(\ref{3.5}), it was shown in \cite{Ku15} that
\begin{equation}\label{3.6}
 G_{3,3}^{3,2} \Big (   {5/2-j,5/2-j,2 \atop 1, 1+k,\dots, 1+k} \Big |1 \Big ) =
 \pi^2 {\Gamma(k) \Gamma^2(2j+2k-1) \over \Gamma^2(j+k)}
 \sum_{\mu=0}^{k-1} {16^{2 - \mu - 2j - k} \Gamma^2(2\mu + 2j - 1) \over
 \Gamma(\mu+1) \Gamma^2(\mu+j) \Gamma(\mu + 2j + k - 1)}.
 \end{equation}
This indeed displays the arithmetic property conjectured in relation to the entries of (\ref{11a}) for $m=2$.

\subsection{Products of truncated orthogonal random matrices}
The random matrix ensemble obtained by considering the leading $N \times N$ sub-block of 
an $(L+N) \times (L+N)$ Haar distributed real orthogonal matrices has been exhibited to have special
integrability properties \cite{KSZ10,Fo10a}. In subsequent works by Kumar and collaborators
(including the present author)
\cite{FK18,FIK20} this was shown to similarly be true of products of such matrices, in which the
parameter $L$ may vary in the product. Specifically, with $N$ even for definiteness, it was shown
that
\begin{equation}\label{3.7}
p_{m,N}  = \prod_{i=1}^m \prod_{s=0}^{N-1}
{\Gamma((L_i + 1 + s)/2) \over \Gamma((s+1)/2) }  \det [\alpha_{2j-1,2k} ]_{j,k=1,\dots,N/2},
 \end{equation}
 where
 \begin{equation}\label{3.7a}
\alpha_{2j-1,2k} 
  = G^{m+1,m}_{2m+1,2m+1} \Big ( {3/2-j,\dots,3/2-j ;1,L_1/2+k,L_2/2+k,\dots,L_m/2+k \atop
 0,k,\dots,k;3/2-j-L_1/2,\dots,3/2-j-L_m/2} \Big | 1 \Big ). 
  \end{equation}
 Using the recurrence (\ref{3.3}) and associated boundary conditions (\ref{3.4}), (\ref{3.5}), 
 a finite sum result was obtained for
 \begin{equation}\label{3.8}
 G^{3,2}_{5,5} \Big ( {3/2-j,3/2-j ;1,\mu+k,\nu+k \atop 0,k,k;3/2-j-\mu,3/2-j-\nu} \Big | 1 \Big),
  \end{equation}
where all parameters are assumed to be positive integers, as relevant to the case $m=2$ with $L_1,L_2$ even. In particular, this was shown to always equal a rational number,
with all terms in the sum rational.
In fact the recursion  (\ref{3.3}) can be used to establish that for any $m$, with $L_1, \dots, L_m$ even, $p_{m,N}$ is a rational number. Specific examples with $N=m=2$, using the notation
$p_{2,2} = p_{2,2}(L_1,L_2)$ to make explicit the values of $L_1, L_2$ tabulated included
 \begin{equation}\label{3.9}
 p_{2,2}(2,2) =  {20 \over 27} \approx 0.7407, \quad  p_{2,2}(4,4) =  \frac{97984}{128625} \approx 0.7617 ,  \quad  p_{2,2}(4,6) = \frac{649984}{848925} \approx 0.7656.
  \end{equation}
 It is easy to establish at the level of the joint distribution that in the limit $L \to \infty$, the $ N \times N$ sub-block of 
an $(L+N) \times (L+N)$ Haar distributed orthogonal random matrices, upon multiplying by $\sqrt{L}$ limits to a standard
Gaussian random matrices. It follows that as the $L_i$'s increase, the probability $p_{m,N} $ must approach that as for
products of standard
Gaussian random matrices. Taking this limit in  (\ref{3.7}), it was shown in \cite{FK16} that indeed (\ref{11a}) results. For the product of
two independent $2 \times 2$ real Gaussian matrices, one has from \cite{Fo13,Ku15} that the probability of all eigenvalues being real equals
${\pi \over 4} \approx 0.7853$, which is in keeping with the trend seen in (\ref{3.9}).

When the product involves odd $L_i$, the arithmetic structure is more complex. For example, it was demonstrated in \cite{FK18} that 
$p_{2,2}(1,2) = (2 \mathcal G + 5)/(3 \pi)$, $p_{2,4}(1,2) = 
 (29412\mathcal{G}^2+10612\mathcal{G}-6767)/(25200\pi^2)$, where   $\mathcal{G}\approx 0.9159$ is Catalan's constant. 
 
 \section{Particular biorthogonal and Pfaffian ensembles}
 \subsection{The difference Wishart ensemble}
 The structure of (\ref{W1a}) suggests the problem of the statistical state formed by $A_1-A_2$ for $A_1, A_2$ positive definite random matrices.
 When $A_1,A_2$ are density matrices drawn from different Hilbert-Schmidt measures, this problem was first considered in \cite{MZB17}.
 Kumar and Charan \cite{KC20}, for $A_1,A_2$ independent complex Wishart matrices, considered the generalised difference
 $\alpha_1 A_1 - \alpha_2 A_2$, with $\alpha_1,\alpha_2$ scalars.
 
 Two formalisms were applied. 
 With $X,Y$ denoting  $N \times N$ Hermitian matrices with eigenvalues $\{ x_j \}$, $\{y_j\}$,
 the first was based on the matrix integral over Haar unitary matrices \cite{KKS15}
  \begin{equation}\label{7.1}
  \int_{X + U^\dagger Y U > 0} \det (X + U^\dagger Y U )^p \, d \mu^{\rm H}(U) \propto {1 \over \Delta(\mathbf x)  \Delta(\mathbf y)}
  \det \Big [ (x_j - y_k)^{p+N-1} \mathbbm 1_{x_j - y_k > 0} \Big ]_{j,k=1,\dots,N},
   \end{equation}
where with $\mathbf u = (u_1,\dots,u_N)$, $\Delta(\mathbf u) := \prod_{1 \le j < k \le N}(u_k - u_j)$. 
The second made use of what is known as the derivative principle \cite{CDKW14,KZ23,Zh21}
  \begin{equation}\label{7.2}
  f(\mathbf x) = {1 \over \prod_{j=1}^N j!} \Delta(\mathbf x) \Delta(-\partial_{\mathbf x}) f_{\rm diag}(\mathbf x),
  \end{equation}
  where $f(\mathbf x)$ on the LHS refers to the joint eigenvalue PDF of a unitary invariant Hermitian matrix ensemble, 
  $f_{\rm diag}(\mathbf x)$ on the RHS refers to the joint eigenvalue PDF of the diagonal entries, $\Delta(\mathbf x)$ is as
  in (\ref{7.1}) and $\Delta(-\partial_{\mathbf x}) := \prod_{1 \le j < k \le N} (\partial_{x_j} - \partial_{x_k})$.
  
  The two approaches give rise to (different) biorthogonal ensemble structures for the joint eigenvalue PDF. In the case of the use of the
  derivative principle, the joint eigenvalue PDF exhibits the additional structure of a polynomial ensemble
  \cite{KS14,Ku16}. This permitted the general $k$-point correlation function to be expressed in terms of a certain correlation kernel (albeit
  which itself involves $N \times N$ determinants). Independent of these structures, but based instead on methods from free probability theory
  \cite{RE08}, an exact functional form of the global density, involving cube and square roots, was obtained.
  
  \subsection{Combinations of complex Wishart and GUE matrices}
Let $W$ be a complex Wishart matrix (Laguerre unitary ensemble --- LUE) with Laguerre parameter $a=n-N$, and let $G$ be an $N \times N$ matrix from the Gaussian unitary ensemble (GUE). In \cite{Ku15a}
Kumar considered the eigenvalues for the unitary invariant ensemble specified by
\begin{equation}\label{z1}
\alpha_1 W + \alpha_2 G, \qquad \alpha_1,\alpha_2 >0.
\end{equation}
A matrix integral approach was used to obtain an explicit polynomial ensemble form for the joint eigenvalue PDF, i.e.~having the form
\begin{equation}\label{z2}
\prod_{1 \le j < k \le N}(x_k - x_j) \det [ f_{j-1}(x_k) ]_{j,k=1,\dots,N}
\end{equation}
for some $\{ f_{j-1} \}$.
The subsequent work \cite{KR19} (see also \cite{Ki19}) highlights that for all random matrix sums $X+Y$, where $X,Y$ are independent polynomial ensembles, there is a systematic approach using the theory of spherical transforms and spherical functions as introduced into random matrix theory (in the context of unitary invariant products) in \cite{KK16,KK19}. In particular, when one member of the sum is a GUE matrix, scaled so that the diagonal entries are standard 
Gaussians, and the other a general polynomial ensemble (\ref{z2}) (scaled so that the diagonal entries are standard Gaussians) the sum remains a polynomial ensemble, specifically with
$$
f_{k-1}(x) = \int_{-\infty}^\infty e^{-y^2/2} f_{k-1}(x-y) \, dy.
$$

Also considered by Kumar in \cite{Ku15a} were the eigenvalues of  the product $WG$. 
After decomposing $W = X^\dagger X$ for $X$ an $n \times N$ ($n \ge N$)
rectangular standard complex Gaussian matrix, the problem becomes one of studying the eigenvalues of $X^\dagger G X$. Now replace $G$ by any polynomial ensemble (\ref{z2}). It was shown in \cite{FIL18} that the joint eigenvalue PDF of this random product is proportional to 
\begin{equation}\label{z3}
\prod_{1 \le j < k \le N}(x_k - x_j) \det \Big [ 
\int_0^\infty a^{n-N-1} f_{j-1}(x_j/a) \, da  \Big ]_{j,k=1,\dots,N}
\end{equation}
An important intermediate step in establishing (\ref{z3}) is to determine the eigenvalue PDF for the random product $X^\dagger A X$, where $A$ is a fixed Hermitian matrix with a prescribed number of negative eigenvalues. For this purpose, one of the methods given in \cite{FIL18} was to make use of a pseudo-unitary group generalisation of the celebrated
Harish-Chandra/Itzykson-Zuber matrix integral \cite{Fy02,FS02}
\begin{equation}\label{z4}
\int_{U(\eta)/U(1)^N}
e^{- {\rm Tr} \, A V B V^{-1}} (V^{-1} d V)
\propto
{\det [ e^{-a_i b_j} ]_{i,j=1}^{n_0}
\det [ e^{- a_{i+n_0} b_{j+n_0}} ]_{i,j=1}^{N-n_0} \over\prod_{1 \le i < j \le N} (a_j - a_i) (b_j - b_i)},
\end{equation}
where the $a_i$ ($b_i$) are the ordered eigenvalues of $A$ ($B$), with exactly $n_0$ of each negative. Here, with $\eta = {\rm diag} ((-1)^{n_0}, (1)^{N-n_0})$ (here the notation $(a)^k$ denotes $a$ repeated $k$ times), $U(\eta)$ denotes the set of $N \times N$ pseudo unitary matrices specified by the requirement that $V^\dagger \eta V = \eta$.

\subsection{Pandey--Mehta crossover ensemble}
Introduce a weighted sum of $N \times N$ real symmetric matrices from the Gaussian orthogonal ensemble (GOE; see \cite[\S 1.2]{Fo10}), and a complex Hermitian matrix from the GUE according to 
\begin{equation}\label{z5}
\sqrt{1 - \alpha \over 2} A + \alpha B, \quad0 \le \alpha \le 1.
\end{equation}
In \cite{PM83,MP83} the joint eigenvalue
PDF was computed as proportional to
\begin{equation}\label{z6}
e^{- \sum_{l=1}^N \lambda_l^2} 
\prod_{1 \le j < k \le N} ( \lambda_k - \lambda_j)
\, {\rm Pf} \bigg [ \sqrt{1 - \alpha^2 \over 2 \alpha^2}
(\lambda_k - \lambda_j) \bigg ]_{j,k=1,\dots,N},
\end{equation}
valid for $N$ even, while for $N$ odd the size of the determinant must be increased by one, with the final column all entries 1 except the last which is 0, and the final row all entries $-1$ except for the last which must be zero to match the final column.

In \cite{SKK20} Kumar and his students took up the task of computing the distribution of the ratio statistic $r=(\lambda_3 - \lambda_2)/ (\lambda_2 - \lambda_1)$ in the case $N=3$. For the $N=3$ GOE and GUE ensembles the exact evaluation of this statistic (with corresponding PDF
$p_{N}(r) |_{N=3}$ say) has been carried out earlier in \cite{ABGR13}, and shown to be an accurate approximation to the $N \to \infty$ limit of this distribution (in relation to the latter, see the recent work \cite{Ni24}). These results were shown to be special cases of the crossover ensemble
result
\begin{multline}\label{z7}
p_{N}(r) \Big |_{N=3} = {r (r + 1) \over 16 \sqrt{6} \pi (1 - \alpha^2)^{3/2}} \bigg ( {b (5 a^2 + 3 b^2) \over a^4 (a^2 + b^2)^2} + {b r (5 a^2 + 3 b^2 r^2) \over a^4 (a^2 + b^2 r^2)^2} \\
- {b (r+1) (5 a^2 + 3 b^2 (r+1)^2 ) \over a^4
(a^2 + b^2 (r + 1)^2)^2} +
{3 \over a^5} {\rm Arctan} \Big (
{b^3 r ( r + 1) \over a^3 + a b^2 (r^2 + r + 1)}. \Big ) \bigg ),
\end{multline}
where
$a:= ((r^2+r+1)/6)^{1/2}$, $b:=\sqrt{(1-\alpha^2)/8 \alpha^2}$
(take $\alpha \to 0$ for the GOE result, and $\alpha \to 1$ for the GUE).
Starting from this form, it is shown in \cite{SKK20} that the general fractional moment $\int_0^\infty r^q p_{N=3}(r) \, dr$ can be evaluated as a single integral involving the Tricomi function (confluent hypergeometric function of the second kind).

Another study by Kumar and his students \cite{SDK24} relating to the crossover ensemble (\ref{z5}) sought to quantify the multifractal dimension of the eigenvectors. Required for this purpose was the exact distribution of a generic eigenvector component. The latter (PDF $p(x;\epsilon)$ say, with $\epsilon := N \alpha^2$) was given exactly in
\cite{SI94} as a certain double integral. In \cite{SDK24} it was shown that one of the integral over one of the variables therein can be computed exactly, with the simplified exact result reading
\begin{equation}\label{z8}
p(x;\epsilon) = {\epsilon e^{\epsilon} \over 2 \sqrt{\pi}}
\int_0^\pi {\exp(-(\epsilon + 2 x \sin^2(\phi/2)) \csc^2 \phi) \over ( \epsilon + 2 x \sin^2(\phi/2))^{3/2}} 
\Big ( 2 ( \epsilon + 2 x \sin^2(\phi/2)) \csc^2 \phi + 1 \Big )
\, d \phi.
\end{equation}

\section{Quantum conductance statistic}
\subsection{Formalism}
The quantum conductance problem relates to a mesoscopic wire of length
much smaller than the coherence length, but much larger than the mean free path, as distinguishes the metallic phase \cite{MK04}. At each end the wire is connected to electron reservoirs of different chemical potentials, creating a current. The quantity of interest is the dimensionless conductance $G/G_0$, where $G_0 = 2 e^2/h$ is twice the fundamental quantum unit of conductance.

It is assumed that at the left (right) end the wire permits $n$ ($N$) channels, with $n \ge N$ for convenience (channels are the plane wave states distinguished by their wavenumber). Associated with these states are the $2n$ components vector $(\vec{I} , \vec{O} )$ (left end) and the $2m$ component vector $( \vec{I}' , \vec{O}' )$ (right end). Here $\vec{I}$ 
($\vec{O}$) and $\vec{I}'$ 
($\vec{O}'$) denote the $n$ ($N$) component amplitude of the plane wave states travelling into (out of) the left and right ends of the wire. Flux conservation requires that $|\vec{I}|^2 + |\vec{O}|^2 = 1$. The $(n+N) \times (n+N)$ size unitary matrix $S$ relating the flux travelling into the wire from either end, to the flux travelling out according to
$$
S \begin{bmatrix} \vec{I} \\ \vec{I}' \end{bmatrix}
= \begin{bmatrix} \vec{O} \\ \vec{O}' \end{bmatrix}
$$
is referred to as the scattering matrix.
Further, it is convenient to decompose $S$ into blocks relating to the reflection and transmission of the current according to
$$
S = \begin{bmatrix} r_{ n \times n} & t_{n \times N}' \\ t_{ N \times n} & t_{N \times N}\end{bmatrix}.
$$
A key formula --- referred to as the Landauer formula --- gives for the dimensionaless conductance (see e.g.~\cite{Be97})
\begin{equation}\label{G0}
G/G_0 = {\rm Tr} \, t^\dagger t = \sum_{j=1}^N \lambda_j,
\end{equation}
where in the second equality $\{\lambda_j\}$ refers to the eigenvalues of $t^\dagger $.

In the case of perfect leads between the wire and the reservoirs it is hypothesised that $S$ is well described by an $(n+N) \times (n+N)$ random Haar distributed unitary. Hence, according to (\ref{G0}), the question becomes to specify the distribution of the squared singular values of an $N \times N$ sub-block on the such a Haar unitary. It was deduced in \cite{MPK88,Be97} that this is specified by the PDF proportional to
\begin{equation}\label{51}
\prod_{j=1}^N \lambda_j^{\beta a/2} \mathbbm 1_{0 < \lambda_j < 1}
\prod_{1 \le j < k \le N} | \lambda_k - \lambda_j |^{\beta},
\end{equation}
where $a = n - N -2/\beta + 1$ and $\beta = 2$ (if there is no time reversal symmetry the scattering matrix should belong to Dyson's circular orthogonal ensemble of symmetric unitary matrices --- see \cite[\S 2.2.2]{Fo10} --- of size $(n + N) \times (n+N)$ and the PDF for the squared singular values of an $N \times N$ sub-block is given by (\ref{51}) with $\beta = 1$).
Replacing each $ \lambda_j^{\beta a/2}$ by $\lambda_j^{c_1}(1 - \lambda)^{c_2}$ the corresponding PDF specifies the Jacobi $\beta$ ensemble with Jacobi parameters $(c_1,c_2)$ --- thus the Jacobi parameters in (\ref{51}) are $(\beta a/2,0)$.

Referring back to (\ref{G0}) one has $G/G_0 = \sum_{j=1}^N \lambda_j$, and so also has the interpretation as the trace statistic for particular Jacobi $\beta$ ensembles. The PDF for the trace statistic is obtained by averaging $\delta( t - \sum_{l=1}^N \lambda_l)$ against the PDF for the Jacobi $\beta$ ensemble.  Taking the Laplace transform of this average (using the scaled Laplace variable $\beta s/2$ for later convenience) gives now the Jacobi $\beta$ ensemble average of $e^{-\beta s \sum_{j=1}^N \lambda_j/2}$. In the case of Jacobi parameters $(\beta a/2,0)$,
changing variables $s \lambda_j \mapsto \lambda_j$ shows
that up to normalisation this new average can be written
\begin{equation}\label{52}
s^{-(\beta a/2 + 1) N -\beta N (N-1)/2} Q_N(s), \quad Q_N(s) = \int_0^s d \lambda_1 \cdots \int_0^s d \lambda_N \, \prod_{l=1}^N \lambda_l^{\beta a/2} e^{-\beta \lambda_l/2}
\prod_{1 \le j <k \le N} | \lambda_k - \lambda_j|^\beta.
\end{equation}

The multiple integral $Q_N(s)$ has an interpretation in random matrix theory quite distinct from the trace statistic of a class of Jacobi $\beta$ ensembles. Thus, up to normalisation, it specifies
the probability that there are no eigenvalues in the interval $(s,\infty)$ of the Laguerre $\beta$ ensemble (specific Laguerre weight $ \lambda^{\beta a/2} e^{-\beta \lambda/2} \mathbbm 1_{ \lambda > 0}$). The derivative of this probability with respect to $s$ gives the distribution of the largest eigenvalue. In the case $\beta = 2$, the distribution of the smallest eigenvalue
in this same ensemble occurred in the context of the Hilbert-Schmidt measure. 
One recalls from \S \ref{S1.3} that there the special structure (\ref{e5a}) for $a$ a non-negative integer facilitated the computation of the required inverse Laplace transform.
To make use of (\ref{52}) for purposes of computing the PDF of the particular Jacobi $\beta$ ensemble trace statistic (or equivalently the dimensionless conductance), it is also
necessary to be able to express $Q_N(s)$ in a form that permits the
computation of the inverse Laplace transform.

\subsection{Some findings of Kumar}
\subsubsection{Recursive computation}\label{S4.2.1}
In the first of several works on the functional form and computation of the distribution of the ordered eigenvalues for classical ensembles joint with
the present author \cite{FK19,FK22,FK23,FK24,FKS24}, Kumar \cite{FK19} took up this problem. It was observed that
for $\beta a/2 =: \tilde{a}$ a non-negative integer, and $\beta$ a positive integer, $Q_N(s)$ permits
an evaluation that like (\ref{e5a}) only involves exponentials and powers,
\begin{equation}\label{53}
Q_N(s) = \sum_{j=1}^N e^{-j \beta s/2} \sum_{k = \tilde{a}}^{j \tilde{a} + j (N-j) \beta} d_{jk} s^k
\end{equation}
for some coefficients $\{d_{jk}\}$. 
After substitution in (\ref{54}), this indeed then permits the computation of the required inverse Laplace transform to be calculated. Thus with $P_{G/G_0}(t)$ denoting the PDF of the dimensionless conductance, direct calculation then shows
\begin{equation}\label{55}
P_{G/G_0}(t) = K \sum_{j=1}^N \Theta(t-j) \sum_{k=0}^{j (\tilde{a} + (N-j) \beta)} {d_{jk} \over \Gamma(\beta n N/2 - k)}
 \Big ( {2 \over \beta} \Big )^k (t-j)^{\beta n N/2  - k - 1}, 
 \end{equation}
 where $K= \prod_{l=0}^{N-1} \Gamma( {\beta \over 2}(n+l) + 1) / \Gamma({\beta \over 2} l + 1)$
cf.~(\ref{e5b}). 
Underlying (\ref{53}) are the facts that for $\beta$ an even positive integer, the product over pairs in (\ref{52}) expands to a multivariable polynomial, and
similarly for $\beta$ an odd positive integer with the ordering $\lambda_1 > \cdots > \lambda_N \ge 0$, combined with the one-dimensional integral evaluation
\begin{equation}\label{54}
{1 \over p!} \int_0^s x^p e^{-\beta x/2} \, dx =   \Big ( {2 \over \beta} \Big )^{p+1} \bigg ( 1 - e^{- \beta s/2}
\sum_{k=0}^p \Big ( {\beta s \over 2} \Big )^k \bigg ).
\end{equation}
While such direct computation furthermore gives the explicit coefficients $\{d_{jk} \}$ in (\ref{53}), it is not at all efficient, being restricted for practical purposes to small $\beta$,
$\tilde{a}$ and $N$. A key finding in \cite{FK19} is that there is a systematic recursive procedure, well suited to implementation using computer algebra, for the generation of the functional form (\ref{53}) which is far more efficient.

This requires generalising $Q_N(s)$ to the family of integrals
\begin{multline}\label{4.7}
		L_{p,\nu}^{(\alpha)}[ t^{\lambda_1} e^{-\lambda t}](x) := {p! (\nu-p)! \over \nu! } \int_{0}^x dt_1 \cdots \int_{0}^x dt_\nu \,
		\prod_{l=1}^\nu t_l^{\lambda_1} e^{-\lambda t_l} |x-t_l|^{\alpha}
		\\
		\times  \prod_{1\leq j<k \leq \nu} | t_k -t_j|^{2\lambda} e_p(x-t_1,\ldots,x-t_\nu),
	\end{multline}
	where $e_p(t_1,\ldots,t_\nu)$ are the elementary symmetric polynomials. We have that
\begin{equation}\label{57}	
 L_{0,\nu}^{(0)}[ t^{\tilde{a}} e^{-\beta t/2}](s)  = \int_0^s x^{\tilde{a}} e^{-x} L_{0,N-1}^{(\beta)}[ t^{\tilde{a}} e^{-\beta t/2}](x) \, dx, \quad L_{\nu,\nu}^{(\alpha)}[ t^{\lambda_1} e^{-\lambda t}](x) = L_{0,\nu}^{(\alpha + 1)}[ t^{\lambda_1} e^{-\lambda t}](x),
 \end{equation} 
 and moreover $Q_N(s) =  L_{0,N}^{(0)}[ t^{\tilde{a}} e^{-\beta t/2}](s)$. 
 To be able to make use of the second relation in (\ref{57}) requires a recurrence in $p$ for the integrals (\ref{4.7}),
 which starts with knowledge of $L_{0,\nu}^{(\alpha)}$, and permits the successive computation of
 $L_{1,\nu}^{(\alpha)}, \dots, L_{\nu,\nu}^{(\alpha)}$. In fact, using the abbreviated notation
 $L_p(x):=L_{p,\nu}^{(\alpha)}[ t^{\lambda_1} e^{-\lambda t}](x)$,  for this purpose one has available the differential-difference second order recurrence
 \cite{Ku19,FT19}
 \begin{equation}\label{rL}
	\lambda (\nu-p) L_{p+1}(x) = (\lambda(\nu-p)x+B_p)L_p(x) + x{d\over dx}L_p(x) - D_p x L_{p-1}(x),
	\end{equation}
	to be iterated for  $p=0,1,\dots, \nu-1$,
	where
	$$
		B_p = (p-\nu)(\lambda_1 + \alpha + 1 + \lambda(\nu-p-1)), \qquad
		D_p = p(\lambda(\nu-p)+\alpha+1).
	$$
	When the value $p=\nu-1$ is reached and we have available $ L_{\nu}(x)$ for a given $\alpha$, the second relation in (\ref{57}) can be applied. This must be done
	for each $\alpha = 0,\dots, \beta - 1$. Then $\nu$ is incremented by  making use of the first recurrence in (\ref{57}), where for this to be practical 
	we must be able to compute the  integral on the RHS. For this, the integrand having a structure of the form in (\ref{53}) is essential.
	
	As an explicit example, from the implementation of the above recurrence scheme for the computation of $\{ d_{jk} \}$ in (\ref{53}) (there is computer algebra software associated with \cite{FK19} for this purpose), substitution in (\ref{55}) gives that
\begin{multline}\label{57v}		
P_{G/G_0}(t) \Big |_{\substack{ \beta = 1, N=3\\ \tilde{a} = 0 
}} =  
 {3 \over 8} \Big ( t^5 -(t-1)^3 ( 40 - 10(t-1) + (t-1)^2  ) \mathbbm 1_{t>1}  \\
- (t-2)^3 ( 40 + 10 (t-2) + (t-2)^2 )  \mathbbm 1_{t>2} \Big ),
\end{multline}
supported on $0 < t < 3$. Actually the recursive generation of this result served as a test of the method, as it was already obtained in the PhD study of Kumar (joint with his supervisor Pandey) \cite{KP10a}, where the methodology to obtain the coefficients $\{ d_{jk} \}$ in (\ref{53})  relied on the underlying Pfaffian structure for $\beta = 1$.
Exact evaluations of $P_{G/G_0}(t)$ for $\tilde{a}$ a non-negative integer, $N$ small, and $\beta=4$ were also carried out using a different  underlying Pfaffian structure, and for
$\beta = 2$ using the underlying determinant structure; cf.~the text below (\ref{e5b}).

\subsubsection{Matrix differential equation}
For the Jacobi $\beta$ ensemble with weight $x^a(1-x)^b$ the PDF of the trace statistic $P_{\rm Tr}(t)$ say can, for $b$ a non-negative integer, be expressed as
 \begin{multline}\label{10e}
{P}_{\rm Tr}(t)  = {1 \over S_N(a,b,\beta)}\bigg ( \int_0^t dx_1 \cdots \int_0^t dx_N  -  \sum_{p=1}^N  \binom{N}{p} 
\int_1^t dx_1 \cdots \int_1^t dx_p \\ \times \int_0^t dx_{p+1} \cdots \int_0^t dx_N 
\bigg ) \delta \Big ( t - \sum_{l=1}^N x_l \Big )
\prod_{l=1}^N x_l^a (1 - x_l)^b \prod_{1 \le j < k \le N} | x_k - x_j |^\beta.
\end{multline}
Here $S_N(a,b,\beta)$ is the Selberg integral \cite[\S 4.1]{Fo10} which gives the normalisation for the Jacobi $\beta$ ensemble PDF.
The significance of (\ref{10e}) is that due to the delta function constraint, the support of the $p$-th term is $N \ge t \ge p$ and the support
of the integration variable $x_i$ $(i=1,\dots,p$) is $(1,t-p+1)$, and that of $x_j$ $(j=p+1,\dots,N)$ is $(0,t-p)$. Taking this into consideration, and restricting too to
$\beta$ a positive integer, working in
\cite{FK22} shows
 \begin{equation}\label{10hy} 
 {P}_{\rm Tr}(t)   = {1 \over S_N(a,b,\beta)} \sum_{p=0}^{N-1} \xi_p \binom{N}{p}
   K_N(a,b,p,\beta) \chi_{p \le t \le N} (t - p)^{\gamma_p} F_N^{(p)} (t-p),
  \end{equation}
  where $  \xi_p = (-1)^{p (b+1) + \beta p (p - 1)/2}$ and each $F_N^{(p)}(s)$ is analytic in the range $-1 < s < N-p$, normalised so that $F_N^{(p)}(0)=1$.
  For $a$ a non-negative integer $F_N^{(p)}(s)$ is in fact a polynomial in $s$ of degree $ap+b(N-p) + p (N-p)\beta$.
  With $W_{a,\beta,n}$ denoting the normalisation for the Laguerre $\beta$ ensemble with Laguerre weight $x^a e^{-x}$ and $n$ eigenvalues, the explicit value of $ K_N(a,b,p,\beta) $ is given in \cite{FK22} as $W_{b,\beta,p} W_{a,\beta,N-p}/\Gamma(\eta)$, for $\eta$ a particular function quadratic  in $p,N$ and linear in $\beta,b,a$.
  As an illustration, the appropriate specialisation of (\ref{10hy}) was checked to be consistent with (\ref{57v}).
  
  Another finding in \cite{FK22} is that for $a$ a non-negative integer, when each $F_N^{(p)}(s)$ is a polynomial in $s$ normalised to unity at $s=0$, the polynomial can be computed in terms of the Frobenius type series expansion about a particular singular point of the matrix differential equation satisfied by a particular vector solution, with ${P}_{\rm Tr}(t) $
  as a component. To obtain the matrix differential equation,
  one must first note that the differential-difference system satisfied by the Laplace transform of the trace statistic (this is essentially (\ref{rL})) can itself be identified with a first
  order matrix differential equation. For example, in relation to (\ref{rL}) with $\nu = 3$, the latter reads
   \begin{equation}\label{11h}
   x {d \over dx} \begin{bmatrix} L_0(x) \\ L_1(x) \\ L_2(x) \\ L_3(x) \end{bmatrix} =
   \left ( x \begin{bmatrix} - 3 \lambda & 0 & 0 & 0 \\
   D_1 & -2 \lambda & 0 & 0  \\
   0 & D_2 & - \lambda & 0 \\
   0 & 0 & D_3 & 0 \end{bmatrix} +  \begin{bmatrix} -B_0 &  3 \lambda & 0 & 0  \\
  0 & - B_1 & 2 \lambda & 0   \\
   0 & 0 & -B_2&  \lambda  \\
   0 & 0 & 0 & 0 \end{bmatrix}  \right )  \begin{bmatrix} L_0(x) \\ L_1(x) \\ L_2(x) \\ L_3(x) \end{bmatrix}.
   \end{equation}  
  Taking the inverse Laplace transform of this --- a strategy which has its origin in \cite{Da70} --- gives the sought matrix differential equation relating
  to ${P}_{\rm Tr}(t) $.

 \subsubsection{Structure of the conductance statistic for $\beta = 1$ and $\tilde{a}+1/2$ a non-negative integer}
 In the definition of $a$ below (\ref{51}), one sees that with respect to the underlying matrix parameter $n-N$, in the case $\beta = 1$ the exponent $\tilde{a} = \beta a/2$ can take
 on half integer values, starting at $-1/2$. With $p+1/2=: m$ a non-negative integer, the analogue of (\ref{54}) is
 \begin{equation}\label{54d}
{1 \over \Gamma(p+1)} \int_0^s x^p e^{-x} \, dx =   {\rm erf}(\sqrt{s}) - e^{- s}
\sum_{k=0}^{m-1} {s^{1/2+k} \over \Gamma(k+3/2)}
\end{equation}
Now ordering the eigenvalues $\lambda_1 > \cdots > \lambda_N > 0$, expanding the product of differences in (\ref{52}) as a multivariable polynomial, and applying (\ref{54d}) to
compute the integral over $\lambda_N,\dots,\lambda_1$ in order gives a finite sum analogous to (\ref{53}), but with the added complication that each exponential may also
involve a factor involving powers of ${\rm erf}(\sqrt{s/2})$.

Remarkably, computer algebra implementation of the recursive algorithm of \S \ref{S4.2.1} in the case $\beta = 1$ and $\tilde{a}+1/2$ a non-negative integer
carried out in \cite{FK24} indicate a much simpler structure, with all higher powers of ${\rm erf}(\sqrt{s/2})$ cancelling out at each recursive step.
 It is found for low order cases with $N$ odd that
 \begin{equation}\label{B1}
Q_N(s) =
\sum_{l=1}^{(N+1)/2} \Big ( \sqrt{s} e^{-(2l-1)s/2} p_{l,1}^{\rm o}(s) +
 {\rm erf}(\sqrt{s/2}) e^{-(l-1)s}
p_{l,2}^{\rm o}(s) \Big ),
\end{equation}
for  polynomials $p_{l,1}^{\rm o}(s), p_{l,2}^{\rm o}(s)$; specifically $p_{l,2}^{\rm o}(s)|_{l=1}$ is a constant.
Similarly, for $N$ even the computer algebra implementation of the recursive algorithm indicates
\begin{equation}\label{B2}
Q_N(s) =
\sum_{l=1}^{N/2+1} \Big (  e^{-(l-1)s} p_{l,1}^{\rm e}(s) +
 \sqrt{s}\, {\rm erf}(\sqrt{s/2}) e^{-(l-1/2)s}
p_{l,2}^{\rm e}(s) \Big ),
\end{equation}
again for polynomials
$p_{l,1}^{\rm e}(s), p_{l,2}^{\rm e}(s)$  with
$p_{l,1}^{\rm e}(s)|_{l=1}$ a constant.

The significance of the structures (\ref{B1}) and (\ref{B2}) is that the inverse Laplace transform as required for the computation of
of $P_{G/G_0}(t)$ can be carried out explicitly. This would not be the case if there were higher powers of ${\rm erf}(\sqrt{x/2})$ present.
In particular, these findings give an explanation (and alternative derivation) for the earlier results of Kumar in \cite{KP10a} relating to explicit functional forms
in the case $\beta = 1$ and  $\tilde{a}+1/2$ a non-negative integer deduced from the Pfaffian structure, for example
\begin{equation}\label{3.4d}
P_{G/G_0}(t) \Big |_{\substack{\beta = 1,N=3 \\ \tilde{a}= -1/2}} = {6 \over 7} t^{7/2} \mathbbm 1_{0<t<1} + {3 \over 28} \Big ( 
35 t^3 - 175 t^2 + 273 t - 125 - 8 (t - 2)^{5/2} (t + 5) \Theta(t-2) \Big ) \mathbbm 1_{1<t<3}.
\end{equation}

    \section*{Acknowledgements}
 This work has been supported by the Australian Research Council
discovery project grant DP250102552..
 
\small

\providecommand{\bysame}{\leavevmode\hbox to3em{\hrulefill}\thinspace}
\providecommand{\MR}{\relax\ifhmode\unskip\space\fi MR }
\providecommand{\MRhref}[2]{%
  \href{http://www.ams.org/mathscinet-getitem?mr=#1}{#2}
}
\providecommand{\href}[2]{#2}


\begin{thebibliography}{10}


 \bibitem{ATK11}
S.~Adachi, M.~Toda, and H.~Kubotani, \emph{Asymptotic analysis of singular
  values of rectangular complex matrices in the {L}aguerre and fixed trace
  ensembles}, J. Phys. A \textbf{44} (2011), 292002(8pp).
  
  \bibitem{AIK13} 
  G. Akemann, J.R. Ipsen, and M. Kieburg,  Products of rectangular random
  matrices: Singular values and progressive scattering, {\it Phys. Rev. E} 88 (2013), 052118.
  
  
  \bibitem{Ao87}
K.~Aomoto, \emph{Jacobi polynomials associated with {Selberg's} integral}, SIAM
  J. Math. Analysis \textbf{18} (1987), 545--549.
  
  \bibitem{ABGR13}
Y. Atas, E. Bogomolny, O. Giraud, and G. Roux, Phys. Rev. Lett. \emph{Distribution of the ratio of consecutive
level spacings in random matrix ensembles},
\textbf{110} (2011), 084101.

\bibitem{Be97}
C.W.J. Beenakker, \emph{Random-matrix theory of quantum transport}, Rev. Mod. Phys. \textbf{69} (1997), 731--808.


\bibitem{BZ06}
I.~Bengtsson and K.~Zyczkowski, \emph{Geometry of quantum states: an introduction to
quantum entanglement}, 2nd ed., CUP, 2017.


\bibitem{B+22}
E. Bianchi, L. Hackl, M. Kieburg, M.
Rigol, and L. Vidmar,  \emph{Volume-law entanglement entropy of typical pure quantum states}, PRX
Quantum \textbf{3} (2022), 030201.

\bibitem{Bo98}
A. Borodin, 
\emph{Biorthogonal ensembles.} 
Nucl. Phys. B \textbf{536} (1998), 704--732.


 \bibitem{BF25}
   S.-S.~Byun and P.J.~Forrester,   \emph{Progress on the study of the Ginibre ensembles},
  KIAS Springer Series in Mathematics \textbf{3}, Springer, 2025.
  
  
  \bibitem{CKFM18}
  D. Carrillo, S. Kumar, G. Fraidenraich, and L. L. Mendes, \emph{Bit error
probability for MMSE receiver in GFDM systems}, IEEE Commun.
Lett. \textbf{22} (2018), 942--945.




\bibitem{CDKW14}
M.~Christandl, B.~Doran, S.~Kousidis, M.~Walter,
\emph{Eigenvalue distributions of reduced
density matrices},
 Commun. Math. Phys. \textbf{332} (2014), 1--52.

 \bibitem{DJGK24}
S.V.L. Da Silva, L.P. Jim\'rnez, F.D.A. Garc\'ia, S. Kumar, \emph{Performance Analysis of Equal-Gain Combining Receivers Over Composite Multipath/Shadowing Fading Channels},
IEEE Access \textbf{12} (2024), 90726.

  \bibitem{Da70}
  A.W.~Davis, \emph{On the null distribution of the sum of the roots of a multivariate
  beta distribution}, Ann.~Math. Statist. \textbf{41} (1970), 1557--1562.
  
  
  
  \bibitem{Ed88}
A.~Edelman, \emph{Eigenvalues and condition numbers of random matrices}, SIAM
  J. Matrix Anal. Appl. \textbf{9} (1988), 543--560.
  
  
  \bibitem{Ed97}
A.~Edelman, \emph{The probability that a random real {G}aussian matrix has $k$
  real eigenvalues, related distributions, and the circular law}, J.
  Multivariate. Anal. \textbf{60} (1997), 203--232.



\bibitem{ES20}
C.~Eltschka and J.~Siewert,
\emph{Joint Schmidt-type decomposition for two bipartite pure quantum states}, Phys. Rev. A  \textbf{101} (2020), 022302.


    \bibitem{Fo10}
P.J. Forrester, \emph{Log-gases and random matrices}, Princeton University Press,
  Princeton, NJ, 2010.
  
 \bibitem{Fo10a} 
  P.J. Forrester, \emph{The limiting Kac random polynomial and
truncated random orthogonal matrices}, J. Stat. Mech. \textbf{12},
 (2010) P12018.
  
  \bibitem{Fo13}
P.J. Forrester, \emph{Probability of all eigenvalues real for products of standard
  Gaussian matrices}, J.~Phys. A \textbf{47} (2014), 065202.
  
  
  \bibitem{Fo23}
P.J. Forrester, \emph{A review of exact results for fluctuation formulas in random matrix theory},
Probab. Surveys \textbf{20} (2023), 170--225. 

    \bibitem{FH94}
P.J. Forrester and T.D. Hughes, \emph{Complex {Wishart} matrices and
  conductance in mesoscopic systems: exact results}, J. Math. Phys. \textbf{35}
  (1994), 6736--6747.

  \bibitem{FIK20}  
  P.J.~Forrester, J.R.~Ipsen and S.~Kumar, \emph{How many eigenvalues of a product of truncated orthogonal matrices are real?}
\emph{Exper. Math.}  \textbf{29} (2020), 276--290.

\bibitem{FIL18}
P.J.~Forrester, I.R.~Ipsen and D.-Z.~Liu,
\emph{Matrix product ensembles of Hermite type and the
Hyperbolic Harish-Chandr-Itzykson-Zuber integral}, Ann. Henri Poincar\'e \textbf{19} (2018), 1307.

  
  \bibitem{FK16}  
  P.J. Forrester and M.~Kieburg,  \emph{Relating the Bures measure to the Cauchy
two-matrix model}, {\it Comm. Math. Phys.} 342 (2016), 151--187.

\bibitem{FK18}
P.J. Forrester and S. Kumar, \emph{The probability that all eigenvalues are real for products of truncated real
orthogonal random matrices}, J. Theoret. Probab.  \textbf{31} (2018), 2056--2071.

\bibitem{FK19}
P.J. Forrester and S. Kumar, \emph{Recursion scheme for the largest $\beta$-Wishart-Laguerre
eigenvalue and Landauer conductance in quantum transport}, J. Phys. A \textbf{52} (2019),
42LT02.

\bibitem{FK22}
P.J. Forrester and S. Kumar,
\emph{Differential recurrences for the distribution of the trace of the $\beta$-Jacobi ensemble},
Physica D, \textbf{434} (2022), 133220.

\bibitem{FK23}
P.J. Forrester and S. Kumar,
\emph{Computable structural formulas for the distribution of the $\beta$-Jacobi edge eigenvalues}, Ramanujan J.
\textbf{61}, 87--110
(2023).

\bibitem{FK24}
P.J. Forrester and S. Kumar, \emph{Computation of marginal eigenvalue distributions in the
Laguerre and Jacobi $\beta$ ensembles}, arXiv:2402.16069.

\bibitem{FKS24}
P.J. Forrester, S. Kumar and B.-J.~Shen,
\emph{Computing marginal eigenvalue distributions for the Gaussian and Laguerre orthogonal ensembles},
 arXiv:2411.15635.


\bibitem{FM11}
P.J. Forrester and A.~Mays, \emph{Pfaffian point processes for the {G}aussian
  real generalised eigenvalue problem}, Prob. Theory and Rel. Fields
  \textbf{154} (2012), 1--47.
  
  
 \bibitem{FT19} P.J.~Forrester and A. K.~Trinh,
\emph{Finite size corrections at the hard edge for the Laguerre $\beta$ ensemble},
Stud. Appl. Math. \textbf{143} (2019), 315--336. 
  
  
  \bibitem{Fy02}
Y.V. Fyodorov, \emph{Negative moments of characteristic polynomials of random matrices: Ingham?
Siegel integral as an alternative to Hubbard?Stratonovich transformation}, Nucl. Phys. B \textbf{621}
(2002) 643.

\bibitem{FS02}
Y.V. Fyodorov and E. Strahov, Characteristic polynomials of random Hermitian matrices
and Duistermaat?Heckman localisation on non-compact K?ahler manifolds. Nucl. Phys. B \textbf{630}
(2002) 453.
  
  
  
   \bibitem{Gi07a}  
  O. Giraud, \emph{Purity distribution for bipartite random pure states},  J. Phys. A
  \textbf{40} (2007), F1053.
  
  \bibitem{Ha98}
M.J.W. Hall, \emph{Random quantum correlations and density operator
  distributions}, Phys. Lett. A \textbf{242} (1998), 123.
  
  
  \bibitem{KSZ10}
B.A. Khoruzhenko, H.-J. Sommers, and K.~Zyczkowski, \emph{Truncations of random
  orthogonal matrices}, Phys. Rev. E \textbf{82} (2010), 040106(R) (4pp).
  

\bibitem{Ki19}
M.~Kieburg, \emph{Additive matrix convolutions of P\'olya ensembles and polynomial
ensembles}, Rand. Matr. Th. Appl. \textbf{9} (2019), 2150002.

\bibitem{KK16}
M.~Kieburg and H.~K\"osters, \emph{Exact relation between singular value and eigenvalue
statistics},  Random Matrices: Theory and Applications \textbf{5} (2016), 1650015.

\bibitem{KK19}
M. Kieburg and H. K\"osters, \emph{Products of random matrices from polynomial ensembles},
Ann. Inst. Henri Poincar\'e Probab. Stat. \textbf{55} (2019), 98--126.   
  
  
  
  \bibitem{KKS15}
M. Kieburg, A. B. J. Kuijlaars, and D. Stivigny, 
\emph{Singular value statistics of matrix products with truncated unitary matrices.} 
Int. Math. Res. Not. (2016) 3392.

 \bibitem{KZ23}
M. Kieburg and J.~Zhang,
\emph{Derivative principles for invariant ensembles}, Adv.
Math.  \textbf{413} (2023), 108833.



\bibitem{Ku16}
A. B. J. Kuijlaars,
\emph{Transformations of polynomial ensembles.} 
In ``\emph{Modern Trends in Constructive Function Theory}'' Amer. Math. Soc. (2016) 253.

\bibitem{KR19}
A.B.J.~Kuijlaars and P. Rom\'an, \emph{Spherical functions approach to sums of random Hermitian matrices}, Int. Math. Res. Not., 2\textbf{2019}, 1005--1029.

\bibitem{KS14}
A. B. J. Kuijlaars, and D. Stivigny, 
\emph{Singular values of products of random matrices and polynomial ensembles.} 
Random Matrices: Theor. Appl. \textbf{3} (2014) 1450011.



  
  \bibitem{Ku15}
S.~Kumar, \emph{Exact evaluations of some {M}eijer {G}-functions and
  probability of all eigenvalues real for products of two {G}aussian matrices},
  J. Phys. A, \textbf{48} (2015) 445206.
  
  \bibitem{Ku15a}
S.~Kumar, \emph{Random matrix ensembles involving Gaussian Wigner and Wishart matrices, and
biorthogonal structure}, Phys. Rev. E \textbf{92} (2015), 032903.


  \bibitem{Ku19} S. Kumar, \emph{Recursion for the Smallest Eigenvalue Density of beta-Wishart-Laguerre Ensemble},
  J. Stat. Phys. {\bf 175}, (2019) 126.

  
  \bibitem{Ku20}
  S.~Kumar, \emph{Wishart and random density matrices: Analytical results for the mean-square
Hilbert-Schmidt distance}, Phys. Rev. A \textbf{102} (2020), 012405.

 \bibitem{KC20}
S. Kumar, S.S.~Charan, \emph{Spectral statistics for the difference of two Wishart matrices},
J. Phys. A, Math. Theor. \textbf{53} (2020), 505202.

\bibitem{K+13}
S.~Kumar, A.~Nock, H.-J.~Sommers, T.~Guhr, B.~Dietz, M.~Miski-Oglu, A.~Richter, F.~Sch\"afer,
\emph{Distribution of scattering matrix elements in quantum chaotic scattering}, Phys. Rev. Lett. \textbf{111}
 (2013), 030403.


 \bibitem{KP10}
S.~Kumar and A.~Pandey  \emph{Random matrix model for Nakagami-Hoyt fading}, IEEE Trans. Inf.
Th. \textbf{56} (2010), 2360.

 \bibitem{KP10a}
  S.~Kumar and A.~Pandey, \emph{Conductance distributions in chaotic
  mesoscopic cavities}, J.~Phys.~A \textbf{43} (2010), 285101. 

 \bibitem{KSA17}
S. Kumar, B. Sambasivam, S. Anand, \emph{Smallest eigenvalue density for regular
or fixed-trace complex Wishart-Laguerre ensemble and entanglement in coupled
kicked tops}, J. Phys. A  \textbf{50} (2017), 345201.

\bibitem{LAK21}
A. Laha, A. Aggarwal, S. Kumar, \emph{Random density matrices: analytical results for
mean root fidelity and mean-square Bures distance}, Phys. Rev. A \textbf{104} (2021),
022438.

\bibitem{LK23}
A. Laha, S. Kumar, \emph{Random density matrices: analytical results for mean fidelity
and variance of squared Bures distance}, Phys. Rev. E 107 (2023), 034206.

\bibitem{LK24}
A.~Laha and S.~Kumar, \emph{Random density matrices: Closed form expressions for squared Hilbert-Schmidt distance},
Phys. Rev. A  \textbf{514-515} (2024), 129591.

  
  \bibitem{LW21} 
 S.-H.~ Li and L.~Wei,  \emph{Moments of quantum purity and biorthogonal polynomial recurrence},
  J. Phys. A \textbf{54} (2021), 445204.


\bibitem{LP88}
S. Lloyd and H. Pagels, \emph{Complexity as thermodynamic depth}, Ann. Phys. \textbf{188} (1986),186--213.

\bibitem{Lu78}
E. Lubkin,  \emph{Entropy of an $n$-system from its correlation with a $k$-reservoir}, J. Math. Phys., \textbf{19} (1978), 1028.

\bibitem{Lu69}
Y.~L. Luke, \emph{The special functions and their approximations, {V}ol.~{I}},
  Academic Press, New York-London, 1969.

     \bibitem{Me04}
M.L. Mehta, \emph{Random matrices}, 3rd ed., Elsevier, San Diego, 2004.

\bibitem{MP83}
M.L.~Mehta and A. Pandey \emph{On some Gaussian ensembles of Hermitian Matrices}, J. Phys. A \textbf{16} (1983), 2655.


\bibitem{MZB17}
J.~Mej\'ia, C.~Zapata, A.~Botero,
\emph{The difference between two random mixed quantum states: exact and asymptotic spectral analysis},
 J. Phys. A \textbf{50} (2017), 025301.
 
 \bibitem{MK04}
P.A.~Mello and N.~Kumar N  \emph{Quantum transport in mesoscopic systems}, Oxford
University Press, Oxford, 2004.

\bibitem{MPK88}
P.A.Mello, A.Pereyra,and N.Kumar, \emph{Macroscopic approach to multichannel disordered conductors}, Ann.Phys. \textbf{181} (1988), 290--317.

\bibitem{Ni24}
S.M.~Nishigaki,
\emph{Distributions of consecutive level spacings of
Gaussian unitary ensemble and their ratio: ab
initio derivation},
Prog. Theor. Exp. Phys. 2024 081A01.


 \bibitem{NKSG14}
A.~Nock, S.~Kumar, H.-J.~Sommers, T.~Guhr, \emph{Distributions of off-diagonal scattering matrix
elements: exact results}, Ann. Phys. \textbf{342} (2014), 103.

\bibitem{OSZ10}
V.A. Osipov, H.-J. Sommers, and K.~Zyczkowski, \emph{Random Bures mixed states
  and the distribution of their purity}, J.~Phys. A \textbf{43} (2010), 055302.

  \bibitem{Pa93}
D.~Page,   \emph{Average entropy of a subsystem}, Phys. Rev. Lett.
\textbf{71} (1993),  1291--1294.

\bibitem{PM83}
A.~Pandey and M.L.~Mehta, \emph{Gaussian ensembles of random Hermitian matrices intermediate
between orthogonal and unitary ones}, Commun. Math. Phys. \textbf{87} (1983), 449.

 \bibitem{PPZ16}
Z.~Puchala, L.~Pawela, and K.~Zyczkowski, \emph{Distinguishability of generic quantum states},
Phys. Rev. A  \textbf{93} (2016), 062112.


 \bibitem{RE08}
N.R. Rao and A. Edelman, \emph{The polynomial method for random matrices}, Found.
Comp. Math. \textbf{8} (2008), 649--702.

\bibitem{SDK24}
A. Sarkar, A. Dheer and S. Kumar, \emph{Multifractal dimensions for orthogonal-to-unitary crossover
ensemble}, Chaos \textbf{34}, (2024) 033121.



\bibitem{SK19}
A.~Sarkar and S.~Kumar, \emph{Bures-Hall ensemble: spectral density and average entropies},
J. Phys. A \textbf{52} (2019), 295203.


\bibitem{SKK20}
A. Sarkar, M. Kothiyal, and S. Kumar, \emph{Distribution of
the ratio of two consecutive level spacings in orthogonal
to unitary crossover ensembles}, Phys. Rev. E \textbf{101} (2020), 012216.

\bibitem{SK21}
A. Sarkar and S. Kumar, \emph{Generation of Bures-Hall mixed
states using coupled kicked tops}, Phys. Rev. A  \textbf{103} (2021),
032423.

\bibitem{SK23}
A. Sarkar, S. Kumar, \emph{Entanglement spectrum statistics of a time reversal invariant
spin chain system: insights from random matrix theory,} Eur. Phys. J. B \textbf{96} (2023).
120.

\bibitem{SSK23}
A. Sarkar, S. Sen and  S. Kumar, \emph{Spectral crossovers in non-Hermitian spin chains: comparison
with random matrix theory}, Phys. Rev. E 108 (2023) 054210.

\bibitem{Se96}
S.~Sen, \emph{Average entropy of a quantum subsystem},  Phys. Rev. Lett. \textbf{77} (1996), 1.

\bibitem{SK24}
S. Sen and  S. Kumar, \emph{Exact and asymptotic dissipative spectral form factor for elliptic Ginibre unitary ensemble},
arXiv:2407.17148.

\bibitem{SSK23a}
S.~Sen, H.~Shekhar and S.~Kumar, \emph{Spectral statistics of interpolating random circulant matrix},
Phys. Rev. E \textbf{111} (2025) 034301.


\bibitem{SKK17}
R. K. Singh, Karmeshu, and S. Kumar, \emph{A novel approximation for K distribution: Closed-Form BER using DPSK
modulation in free-space optical communication}, IEEE Photon. J. \textbf{9}  (2017), 7906817.

\bibitem{SI94}
H.-J. Sommers and S. Iida, \emph{Eigenvector statistics in the
crossover region between Gaussian orthogonal and 
unitary ensembles}, Phys. Rev. E. \textbf{49} (1994), R2513. 

 \bibitem{SZ04} 
  H.-J. Sommers and K.~\.Zyczkowski, \emph{Statistical properties of random density matrices},
  J.~Phys.~A {\bf 37} (2004), 8457--8466.
  
  \bibitem{We20a}
  L.~Wei, \emph{Proof of Sarkar-Kumar's conjectures on average entanglement
entropies over the Bures-Hall ensemble}, J. Phys. A  \textbf{53}
(2020), 235203. 

 \bibitem{We20b}
 L.~Wei, \emph{Exact variance of von Neumann entanglement entropy over the Bures-Hall
measure}, Phys. Rev. E \textbf{102} (2020), 062128.
 
  \bibitem{WW23}
  L.~Wei and N.S.~Witte, \emph{Quantum interpolating ensemble: bi-orthogonal polynomials and
average entropies}, Random Matrices: Theory Appl. \textbf{12} (2023), 2250055.
  
  \bibitem{YHOW24}
  L. Ye, Y. Huang, J.C. Osborn, L. Wei, \emph{Square root statistics of density matrices and
their applications}, Entropy \textbf{26} (2024), 68.

\bibitem{Zh21}
J.~Zhang, \emph{Decompositions, invariances and harmonic analysis in random matrix theory},
PhD thesis, University of Melbourne, 2021.
  

\bibitem{ZS01}
K.~Zyczkowski and H.-J. Sommers, \emph{Induced measures in the space of mixed
  quantum states}, J.~Phys. A \textbf{34} (2001), 7111--7125.

\bibitem{ZS03}
K.~Zyczkowski and H.-J. Sommers, \emph{Bures volume of the set of mixed quantum states}, J.~Phys. A
  \textbf{36} (2003), 10083--10100.


\end{thebibliography}
\end{document}